\documentclass{resonance}

\usepackage{amsfonts}
\usepackage{amsmath}
\usepackage{amssymb}
\usepackage{color}
\def\i{\item}

\newcommand{\bed}{\begin{displaymath}}
\newcommand{\eed}{\end{displaymath}}
\newcommand{\bei}{\begin{itemize}}
\newcommand{\eei}{\end{itemize}}
\newcommand{\bef}{\begin{figure}}
\newcommand{\eef}{\end{figure}}
\newcommand{\ben}{\begin{enumerate}}
\newcommand{\een}{\end{enumerate}}
\newcommand{\beq}{\begin{equation}}
\newcommand{\eeq}{\end{equation}}
\newcommand{\ber}{\begin{eqnarray}}
\newcommand{\eer}{\end{eqnarray}}

\newcommand{\gcc}{\mbox{${\rm g} \, {\rm cm}^{-3}$}}

\newcommand{\msun}{\mbox{{\rm M}$_{\odot}$}}

\newcommand{\lsim}{\raisebox{-0.3ex}{\mbox{$\stackrel{<}{_\sim} \,$}}}
\newcommand{\gsim}{\raisebox{-0.3ex}{\mbox{$\stackrel{>}{_\sim} \,$}}}

\newcounter{attnctr} 

\definecolor{cblue}{rgb}{0.9,0.9,1.0}
\definecolor{darkblue}{rgb}{0.1,0.1,0.6}
\definecolor{darkred}{rgb}{0.6,0.1,0.1}
\begin{document}


\title{Gravity Defied \\
     From potato asteroids to magnetised neutron stars}
\secondTitle{IV. Neutron Stars (dead stars of the second kind)}
\author{Sushan Konar}

\maketitle
\authorIntro{\includegraphics[width=2cm,angle=90]{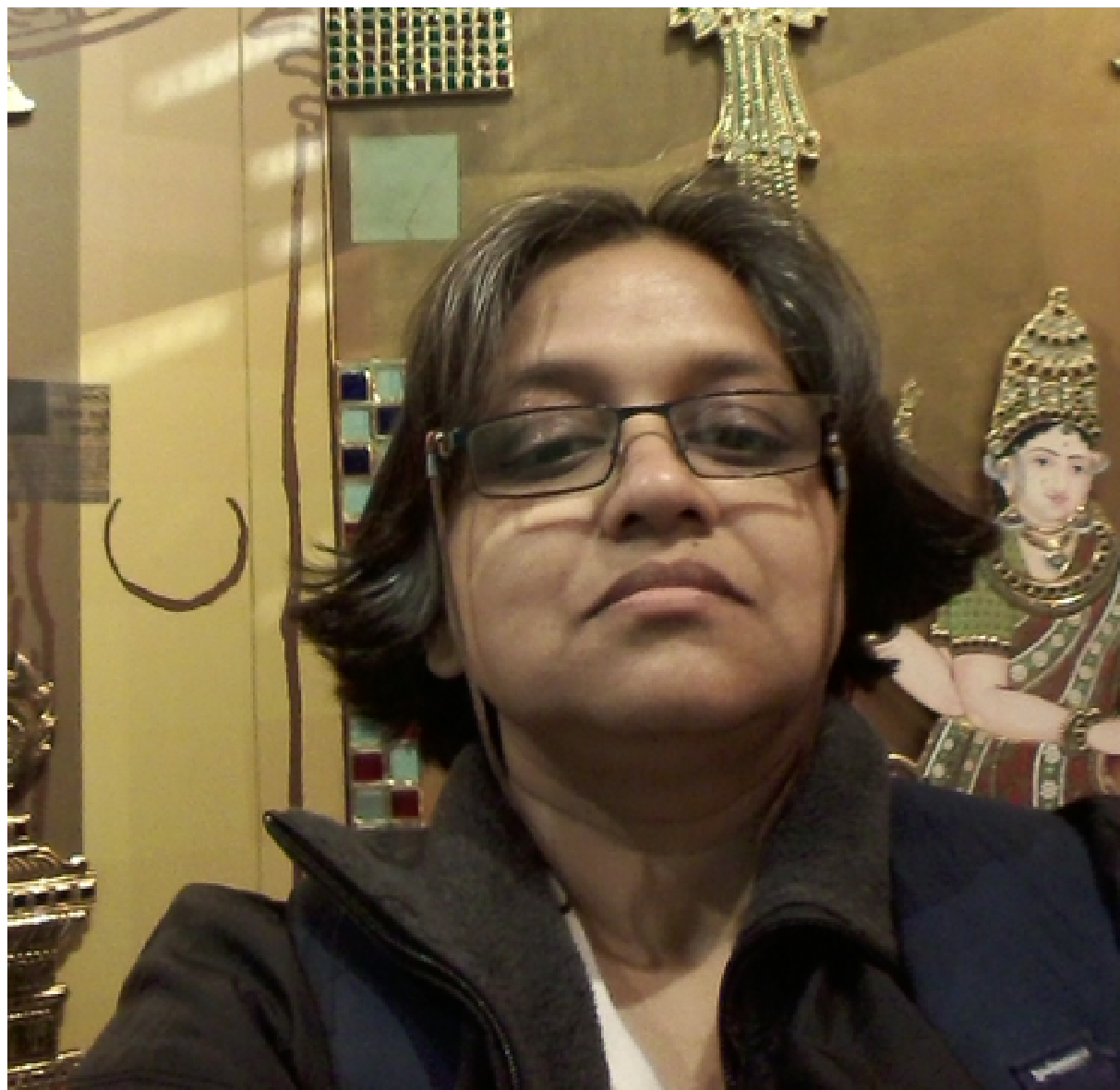}\\
Sushan Konar  works at  NCRA-TIFR, Pune. She  tries to  understand the
physics of stellar  compact objects (white dwarfs,  neutron stars) for
her livelihood and writes a blog about life in academia as a hobby.}
\begin{abstract}
  A  star burns  its  nuclear  fuel and  balances  gravitation by  the
  pressure of the  heated gas, during its active  lifetime.  After the
  exhaustion of  the nuclear fuel,  a low mass  star finds peace  as a
  {\em white dwarf}, where the pressure support against gravitation is
  provided by  Fermi-degenerate electrons. However, for  massive stars
  the gravitational squeeze becomes so  severe that in the final phase
  of evolution,  the average density approximately  equals the nuclear
  density.   At  such  densities  most of  the  protons  combine  with
  electrons  to  convert themselves  into  neutrons.   A {\em  Neutron
    star}, composed  of such  neutron-rich material,  is host  to some
  fascinating physics  arising out of  its amazingly compact  state of
  matter  (where a  solar mass  is packed  inside a  sphere of  radius
  $\sim$ 10Km).
\end{abstract}
\monthyear{2017}
\artNature{GENERAL  ARTICLE}

\section{The beginnings..}

\begin{figure}[!t]
  \caption{Jocelyn Bell, photographed in  front of the Cambridge radio
    telescope in 1967.  Bell was  working with Anthony Hewish, for her
    PhD thesis  on interplanetary scintillation, when  she observed a
    unusually periodic  signal coming from the  sky.  After discarding
    many plausible  and implausible theories (including  one involving
    the `little green  men') these signals were  finally identified to
    be radio emissions coming from a highly magnetised neutron star.}
  \label{f_jocelyn}
\vspace{-0.5cm}
\centering\includegraphics[width=7.5cm]{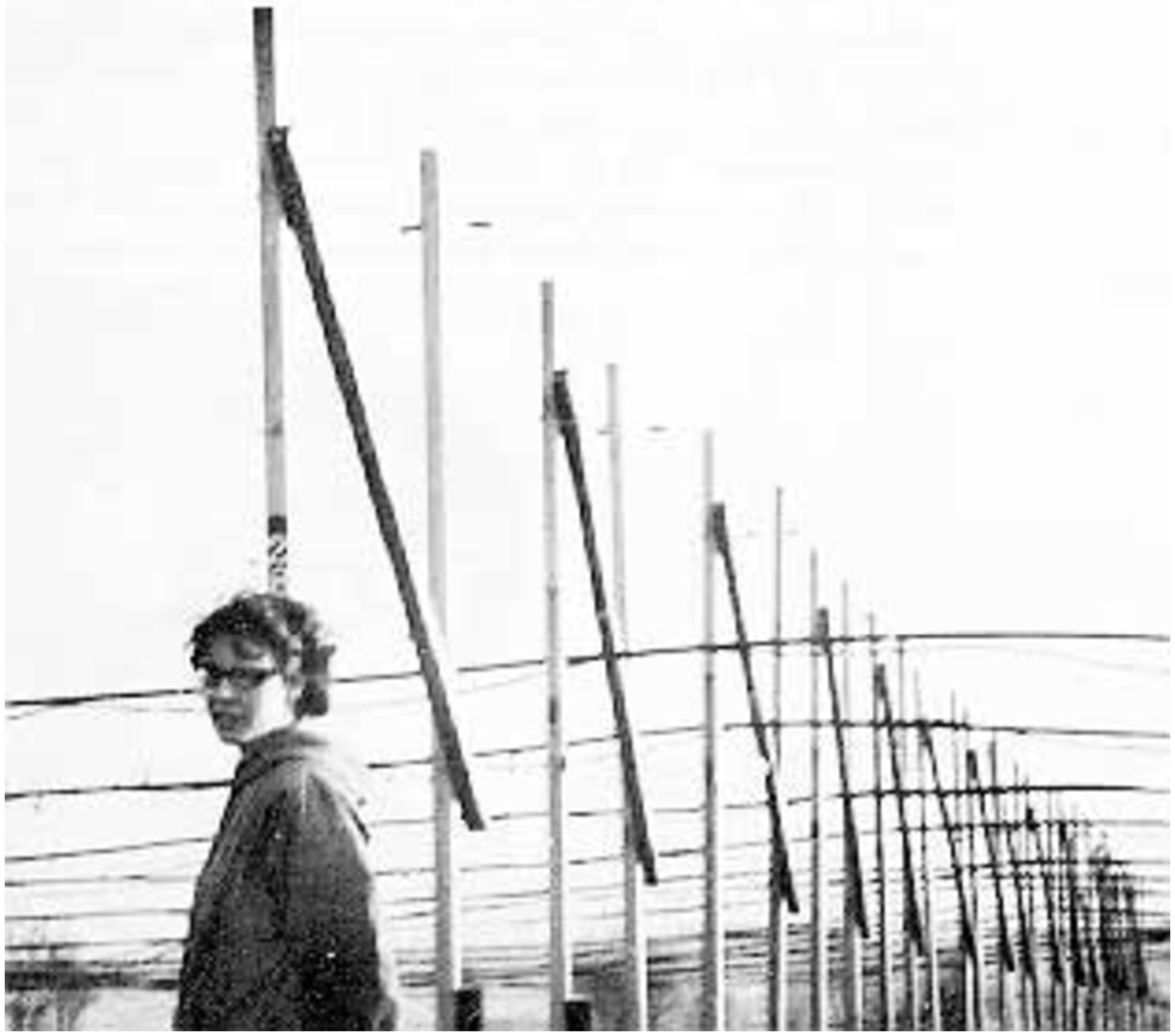}
\end{figure}

The discovery of a radio pulsar by Jocelyn Bell (Fig.~\ref{f_jocelyn})
in 1967 and its subsequent identification  with a neutron star, one of
the most  exotic objects in  the Universe,  is a watershed  moment for
theoretical  astrophysics  when   predictions  made  several  decades
previously were  finally confirmed  through observation. The  story of
neutron  stars  has  always  been  marked  with  prescient  ideas  and
serendipitous  observations,   which  is  not  surprising   given  the
unbelievably rich  and complex  physics these  tiny stars  pack within
themselves.


Along with the understanding of their nuclear energy source, arose the
question about the  end states of stars.  The nuclear  fuel is said to
be exhausted  either when a  star is  unable to reach  the temperature
required  for  the  nuclear  fusion   of  the  next  element  or  when
$_{26}$Fe$^{56}$ is formed ending the fusion reaction chain. In 1930s,
S. Chandrasekhar showed that white dwarfs  could be one such end state
of  stars   where  gravitation   is  balanced   by  the   pressure  of
Fermi-degenerate electrons.  The logical extension of this argument is
the {\em neutron  stars} where the pressure comes  from the degenerate
neutrons and from  the forces of nuclear interaction  making these the
{\em dead stars of the second kind}.


%

Interestingly, this is exactly what  was suggested by Walter Baade and
Fritz Zwicky in 1934, only two years after neutrons were discovered by
James Chadwick  (and almost  three decades  before neutron  stars were
actually  observed)!   The discovery  that  Andromeda  is an  external
galaxy  enabled them  to estimate  the energy  released from  the nova
observed  in   1885  to   be  around   $\sim  10^{52}$~erg~\mfnote{The
  gravitational  energy release  associated  with the  formation of  a
  neutron star is calculated to be,
\beq
E_{\rm   G}  \simeq   G  \frac{M^2}{R}   \simeq  10^{52}~\mbox{erg}\,;
\nonumber
\eeq
where  $M$  =  1\msun~and  $R$  =  10~Km  has  been  assumed.}.   They
hypothesised that  this must  be associated with  the formation  of an
object  of  radius $\sim$10~Km  and  such  an  object must  be  almost
entirely  made  up  of  neutrons.    In  the  mid  1930s,  Lev  Landau
strengthened  this  argument  by  showing that  beyond  a  density  of
$10^{11}$~\gcc, electrons would combine with protons to form neutrons.
To understand how prescient this suggestion was, one needs to remember
that   one  of   the  end   products   of  this   process,  known   as
$\beta$-capture, is a {\em neutrino} which would be discovered only in
1956~\mfnote{$\beta$-capture reaction :
\vspace{-0.25cm}
\beq
p^+ + e^- \rightleftharpoons n + \nu \,. \nonumber
\eeq
}!

In the 1940s, it was pointed out  that the position of the crab nebula
coincided with  that of the 1054  guest star, recorded by  the Chinese
court astronomers.  The  discovery of a radio pulsar at  the centre of
Crab nebula  vindicated Baade and Zwicky's  hypothesis completely. The
current understanding  from stellar  evolution theories is  that stars
with  main  sequence  masses  of  8 -  20  \msun~end  their  lives  in
supernovae.   Most of  the  stellar  mass is  thrown  away  in such  a
supernova explosion and a compact neutron star ($M \sim 1.4 \msun$, $R
\sim 10$~km) is left behind.
 
\section{The interior..}

\begin{figure}
  \caption{The neutron-proton  ratio in terrestrial nucleons  is shown
    with the stability band  (greenish region). Picture courtesy -
    {\em   Chemistry,  Mcmury   Fay,  Prentice-Hall,   2004.}  A   few
    neutron-rich  elements,  expected to  be  found  in neutrons  star
    crusts are marked in magenta.}
  \label{f_nucleon}
\vspace{-0.5cm} 
\centering\includegraphics[width=10.0cm]{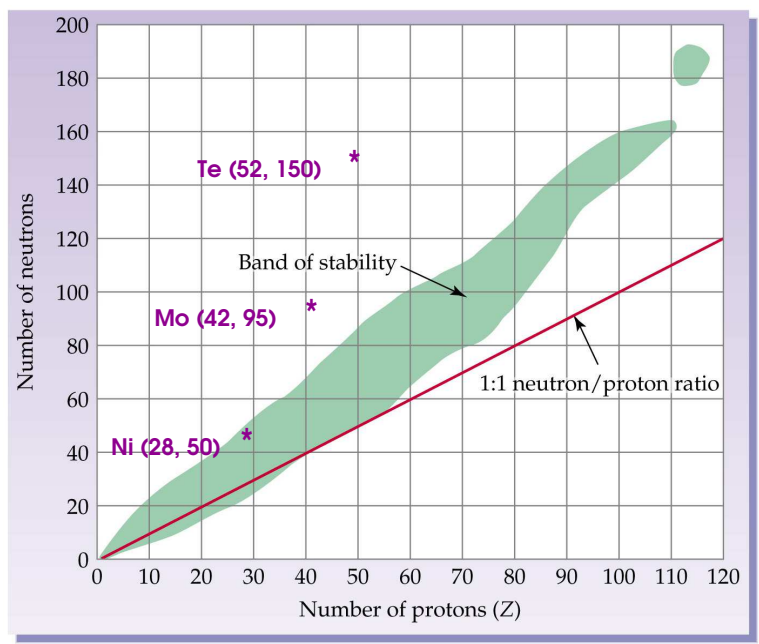}
\vspace{-0.5cm}
\end{figure}

One of the  key theoretical concepts associated with  neutron stars is
the stability of neutrons inside it. Free neutrons decay~\mfnote{Decay
  of  free neutrons  is  known  as $\beta$-decay  and  is the  reverse
  process of $\beta$-capture, where a neutron decays into a proton, an
  electron and an anti-neutrino.
\vspace{-0.25cm}
\beq
n \rightleftharpoons p^+ + e^- + \overline{\nu} \,. \nonumber
\eeq
} with a half-life of $\sim$15  minutes. The forward direction of this
reaction is  prohibited inside  a neutron  star. As  mentioned before,
free neutrons  exist only beyond  a density of $\sim  10^{11}$\gcc. At
such densities,  the electrons  are degenerate  with very  large Fermi
energies ($E_{\rm  F} \propto \rho^{1/3}$, because  electrons are also
relativistic at these densities). Since,  all quantum levels with $E <
E_{\rm  F}$  would  be  occupied,   an  electron  produced  through  a
$\beta$-decay  process would  require  an energy  larger than  $E_{\rm
  F}$. A  neutron at  rest is  unable to supply  such energies  to the
electron produced and the $\beta$-decay is therefore prohibited.

As in the  case of a white  dwarf, the structure of a  neutron star is
obtained  by  solving the  hydrostatic  pressure  balance equation  in
conjunction with the equation of  state. Except, a neutron star being
far more  compact (with  $V_{\rm E}/c  \lsim 0.5$),  we would  have to
solve  the TOV  equation instead  of the  non-relativistic hydrostatic
equation~\mfnote{see  S. Konar,  Gravity Defied,  Resonance, Vol  22,
  No.5}. The main problem though, lies in finding the correct equation
of state, that is identifying the correct state of matter.

\begin{table}
  \caption{Neutron-rich nuclei expected to be  found in the crust of a
    neutron star. The neutron to proton ratio increases with increasing
    density as  more and  more protons  convert into  neutrons through
    $\beta$-capture. Stable  terrestrial counterparts of  these nuclei
    have  also  been  shown  for  comparison.  It  can  be  seen  from
    Fig.~\ref{f_nucleon}  above  that  the nucleons  expected  in  the
    neutron star  crust have  very different  neutron to  proton ratio
    compared to their terrestrial counterparts. The data is taken from
    {\em Baym, Pethick and Sutherland, 1971, ApJ, 170, 299}.}
\begin{tabular}{rllllr}
$\rho$ && Z && A & terrestrial\\
\gcc && && & element \\
$10$              && 26  &&  56  & $^{26}$Fe$_{56}$  \\ 
$10^6$            && 26  &&  56  & $^{26}$Fe$_{56}$  \\ 
$10^7$            && 28  &&  62  & $^{28}$Ni$_{58}$  \\ 
$10^8$            && 28  &&  62  & $^{28}$Ni$_{58}$  \\ 
$10^9$            && 28  &&  64  & $^{28}$Ni$_{58}$  \\ 
$10^{10}$          && 32  &&  82  & $^{32}$Ge$_{72}$  \\ 
$10^{11}$          && 28  &&  78  & $^{28}$Ni$_{58}$  \\ 
$10^{12}$          && 42  &&  137 & $^{42}$Mo$_{96}$  \\ 
$10^{13}$          && 52  &&  200 & $^{52}$Te$_{128}$ \\ 
$5 \times 10^{13}$ && 74  &&  375 & $^{74}$W$_{184}$  \\ 
\end{tabular} 
\label{t_nscrust}
\end{table}
\begin{figure}
  \caption{Composition of a neutron star.  There is a huge uncertainty
    about the state  of matter at highest densities, in  the core of a
    neutron star  (denoted by a question  mark). Different assumptions
    about  this result  in different  mass-radius relations.  Accurate
    mass measurements  can lift the  degeneracy about core composition
    as  can be seen    from   Fig.~\ref{f_eos}. Picture   courtesy    -
    {\em https://heasarc.gsfc.nasa.gov/}}
  \label{f_interior}
\vspace{-0.5cm} 
\centering\includegraphics[width=10.0cm]{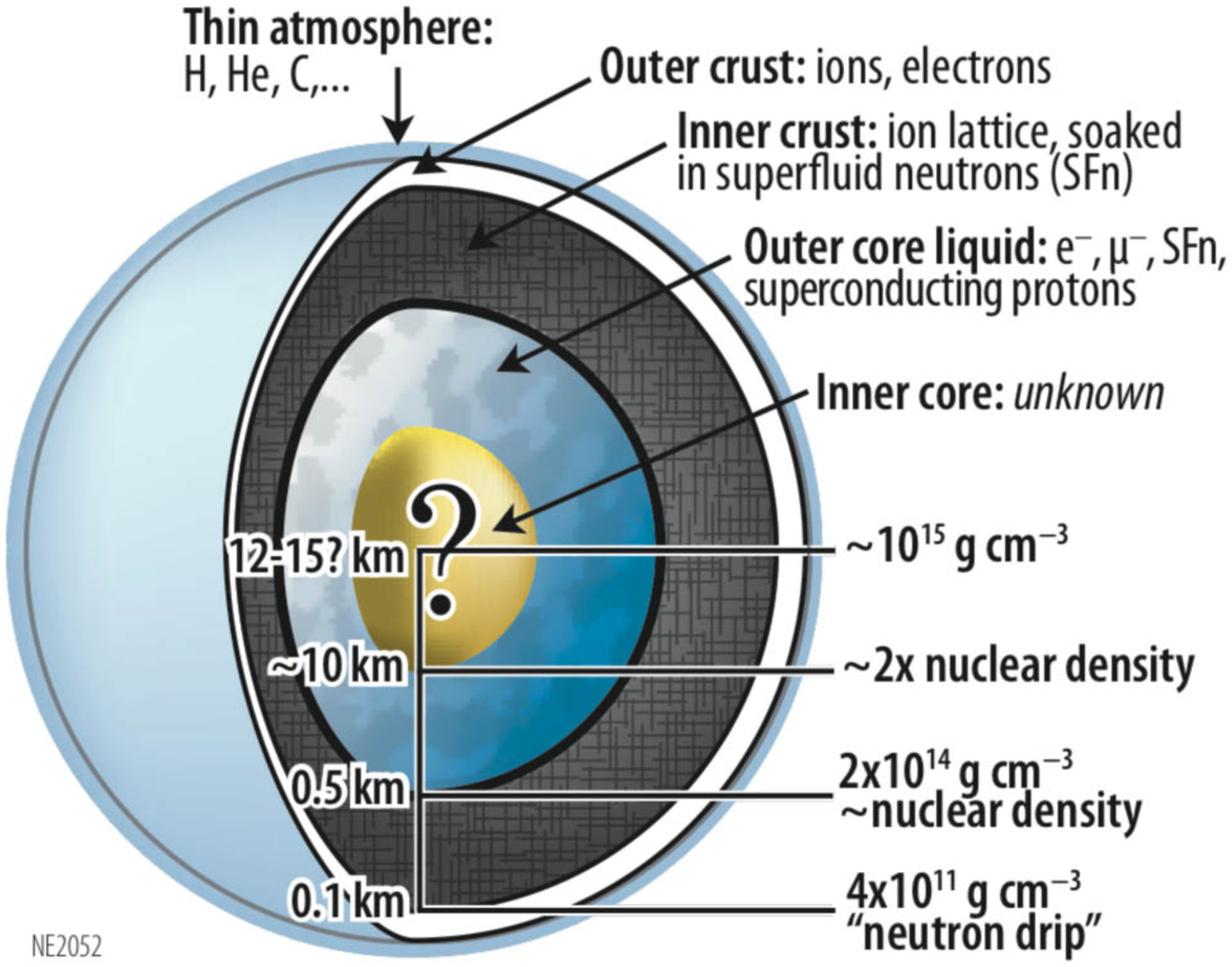}
\vspace{-0.5cm}
\end{figure}

Neutron star matter encompasses a  wide range of densities, from $\sim
10^6$\gcc~at the surface  to several times the nuclear  density in the
stellar  core  (Fig.~\ref{f_interior}).  At  a  given  density,  the
equilibrium  composition  is the  one  which  is most  stable  against
$\beta$-capture. Starting from ordinary  Fe$^{56}$ at the surface, the
equilibrium    nuclide   becomes    more    and   more    neutron-rich
(Table~\ref{t_nscrust}) as the density increases. Though neutron-rich,
the material  is expected to  arrange itself into a  crystalline solid
with the  electrons forming a  Fermi-degenerate gas.  At a  density of
$\sim 4  \times 10^{11}$\gcc~the neutrons  are so numerous  that their
highest energy level, inside a  nucleus, reach the continuum value and
some of the neutrons become free of the nuclear binding (`drip' out of
the nucleus). Therefore, from this `drip' density onwards, the neutron
star material  is composed of  a neutron-rich nuclei (again  forming a
crystalline  solid),  Fermi-degenerate electrons  and  a  gas of  free
neutrons. It is understood that this free neutrons actually exist in a
super-fluid   state,  giving   rise   to  a   number  of   interesting
observational consequences.  The solid crystalline phase  extends upto
the nuclear density and this outer part  of a neutron star is known as
the {\em crust}.

\begin{figure}
  \caption{A schematic diagram showing a radio pulsar undergoing an
  episode of `glitch' - a sudden spin-up.}
  \label{f_glitch}
\vspace{-0.5cm} 
\centering\includegraphics[width=7.5cm]{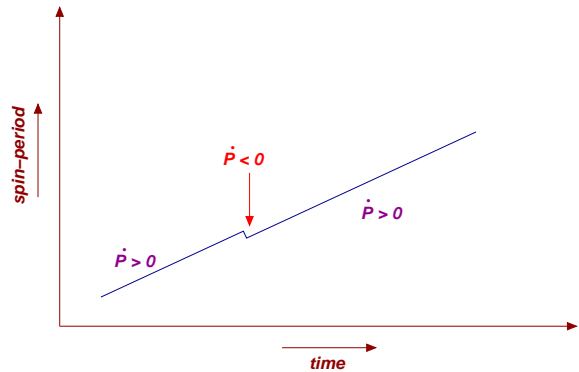}
\vspace{-0.5cm} 
\end{figure}

The  nuclei completely  dissolve into  its constituent  particles when
nuclear density is reached in the inner  part of the star known as the
{\em core}.   It is believed that  both neutrons and protons  exist in
separate super-fluid phases (neutron forming a neutral superfluid, and
protons forming a charged superconductor) inside the core of a neutron
star. It  is also understood  that the rotation  of a neutron  star is
supported  by  creation of  Onsager-Feynman  vortices  in the  neutron
superfluid,  whereas  the magnetic  field  is  supported by  Abrikosov
fluxoids  generated in  the proton  superconductor~\mfnote{Protons are
  thought  to form  a  type-II  superconductor inside  the  core of  a
  neutron star and can support  magnetic fields in quantised Abrikosov
  fluxoids, each of which carry a flux quantum $\phi$ given by
\beq
\phi = \frac{hc}{2e} = 2 \times 10^{-7}~\mbox{G-cm$^2$} \,. \nonumber
\eeq
Alexei Abrikosov (1928-2017) was instrumental in developing the theory
of `type-II superconductors', a phase of matter that later found to be
of great  important for modern  technology.} Both of  these superfluid
vortices are created when neutron star matter goes into the superfluid
phase.  It is understood that the superfluid transition temperature of
the neutron star material is $\gsim  10^9$K and the neutron star cools
down to such temperatures (from $\sim  10^{11}$K at the time of birth)
within a very short time after its formation in a supernova explosion.

It  is difficult  to envisage  the existence  of superfluidity  inside
neutron stars except from  a theoretical viewpoint. Fortunately, there
are indirect evidences to support  this conjecture. Many radio pulsars
have been observed to undergo a  period `glitch'. The radiation from a
radio  pulsar comes  at the  expense of  its rotational  energy. As  a
result  a radio  pulsar  undergoes a  secular  spin-down.  Some  radio
pulsars are  observed to  undergo a  sudden spin-up,  with a  total or
partial    recovery    to    its   original    rate    of    spin-down
(Fig.~\ref{f_glitch}).  We  can understand  the origin of  this sudden
spin-up in  the following way.   The crust  of a neutron  star rotates
like  a rigid  body.   However, the  superfluid  component rotates  by
creating  vortices each  of which  carry  a finite  amount of  angular
momentum. To  slow down,  the superfluid  needs to  expel a  number of
these vortices  pinned to  the crustal  lattice through  the neutron's
interaction with the  nucleons. So, even though the  solid crust slows
down,  the  superfluid  component  continues to  rotate  at  a  faster
rate. As  the difference in the  rotation rate increases, so  does the
tension at the  sites the vortices are pinned. When  the difference in
rotational  energy  overcome the  pinning  energy  there is  a  sudden
un-pinning of the vortices and  they are expelled from the superfluid.
Evidently, this results in the solid  crust gaining an extra amount of
angular  momentum which  shows up  as  a sudden  spin-up.  Though  the
spin-up itself  can be  explained in terms  of the  super-fluids (there
exist other theories that do  not involve super-fluids), it is actually
the  post-glitch  behaviour  of   certain  pulsars  that  definitively
indicates the presence of a superfluid inside a neutron star.

\begin{figure}
  \caption{Inside  the core  of a  neutron star,  when the  density is
    $\simeq 8 \rho_{\rm nuclear}$, the neutrons and protons are expected
    to undergo a deconfinement transition to a quark phase composed of
    `up'  and `down'  quarks. Apparently,  the stable  quark phase  is
    achieved  when some  of  these convert  into  `strange' quarks  and
    achieve a 1:1:1 ratio between all three quark flavors.}
  \label{f_deconf}
\vspace{-0.5cm} 
\centering\includegraphics[width=10.0cm]{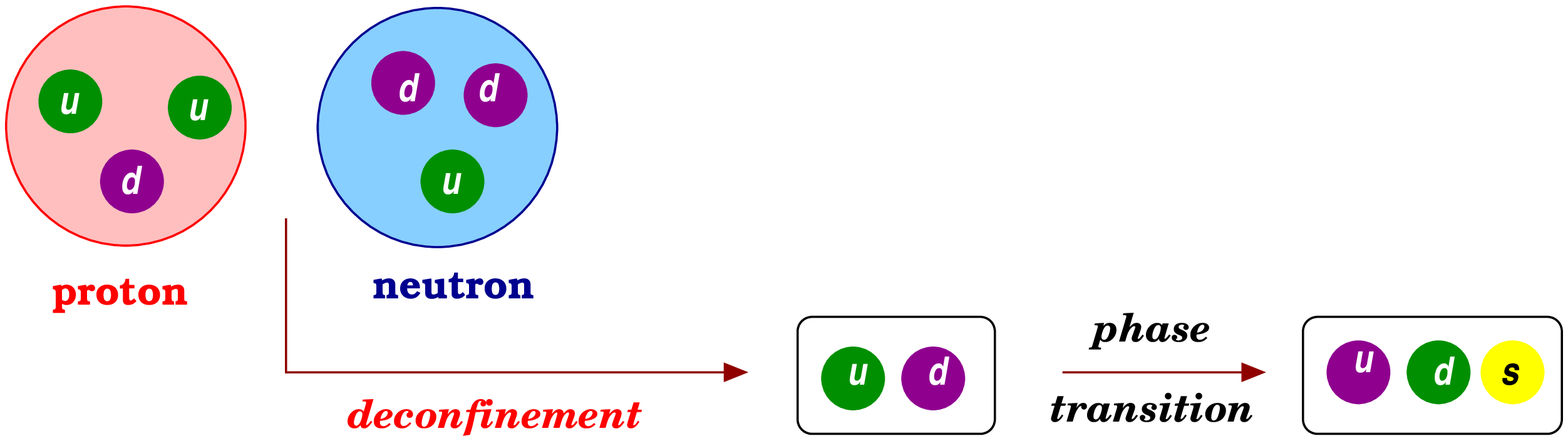}
\end{figure}

\begin{figure}
  \caption{Mass-radius relations for  non-rotating neutron  stars for
    different equations  state (blue - nucleons, pink  - nucleons +
    exotic matter, green - strange quark matter). The horizontal bands
    correspond to  actual observational  mass measurements.  of  a few
    neutron stars.  An EoS line not intersecting an observed mass band
    is ruled out by that measurement. Most exotic matter EoS are ruled
    out by mass measurement of  the neutron star J1614-2230.  The grey
    regions show  parameter space  ruled out  by other  theoretical or
    observational  constraints  (GR -  general  relativity,  P -  spin
    period).  Figure  courtesy -  {\em Demorest  et al.,  2010, Nature,
      467, 1081}}
  \label{f_eos}
\vspace{-0.5cm}
\centering\includegraphics[width=10.0cm]{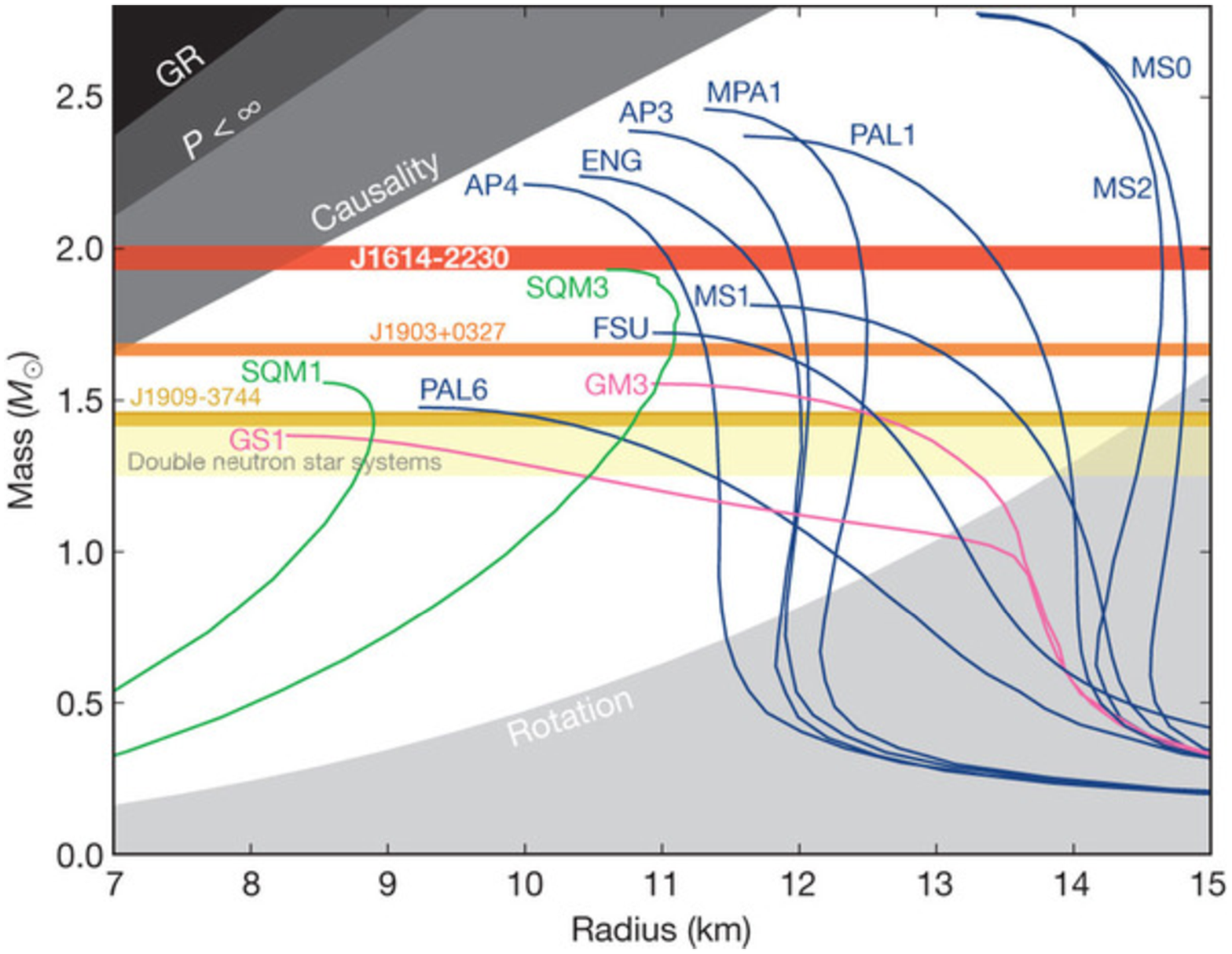}
\vspace{-0.5cm} 
\end{figure}

It  is  conjectured  that  various   novel  phases  like  the  hyperon
matter~\mfnote{A hyperon is  a baryon, containing one  or more quarks.
  These are  highly unstable under terrestrial  conditions and quickly
  decay  into  nucleons. At  densities  $\sim  2 -  3\rho_{\rm  nuc}$,
  realisable in the core of a  neutron star, the inverse reactions can
  take  place  producing   hyperons.},  Bose-Einstein  condensates  of
strange mesons and quark matter can exist inside the core, giving rise
to  a plethora  of  possible  equations of  state.   For example,  EoS
involving the strange  quark matter (SQM), where up,  down and strange
quarks exist in equal numbers (Fig.~\ref{f_deconf}) and is expected to
appear when the  density is approximately 8 times that  of the nuclear
density, have been rather popular  for a while.  Fortunately, observed
masses  and  radii  of  neutron   stars  can  directly  constrain  the
composition and the  EoS of the interior.   Technological advances has
allowed for rather precise mass  measurements in recent years.  It can
be seen  from Fig.~\ref{f_eos}  that mass  measurement of  the neutron
star J1614-2230 has ruled out all EoS based on pure SQM.

\section{The observations..}

It's been half  a century since the  discovery of a neutron  star as a
radio  pulsar and  $\sim$3000 objects  have been  observed during  the
intervening period. As a result of these observations, the spin-period
($P_s$) and the surface magnetic field  ($B_s$) have emerged to be the
most  significant   intrinsic  parameters  of  a   neutron  star.  The
population of neutron  stars is seen to display a  wide range of these
parameters ($P_s \sim 10^{-3} - 10^2$~s, $B_s \sim 10^8 - 10^{15}$~G)
and have diverse observational characteristics. However the population
can  be broadly  classified into  three categories,  depending on  the
nature of the energy generation in these stars.
 
\begin{figure}
  \caption{A neutron star in its  radio pulsar phase. The misalignment
    between its rotation and magnetic axis makes the beam of radiation
    sweep through  the space, giving  it a lighthouse  effect. Picture
    courtesy - {\em http://www.scmp.com/}}
%
%
  \label{f_pulsar}
\vspace{-0.5cm} 
\centering\includegraphics[width=5.0cm]{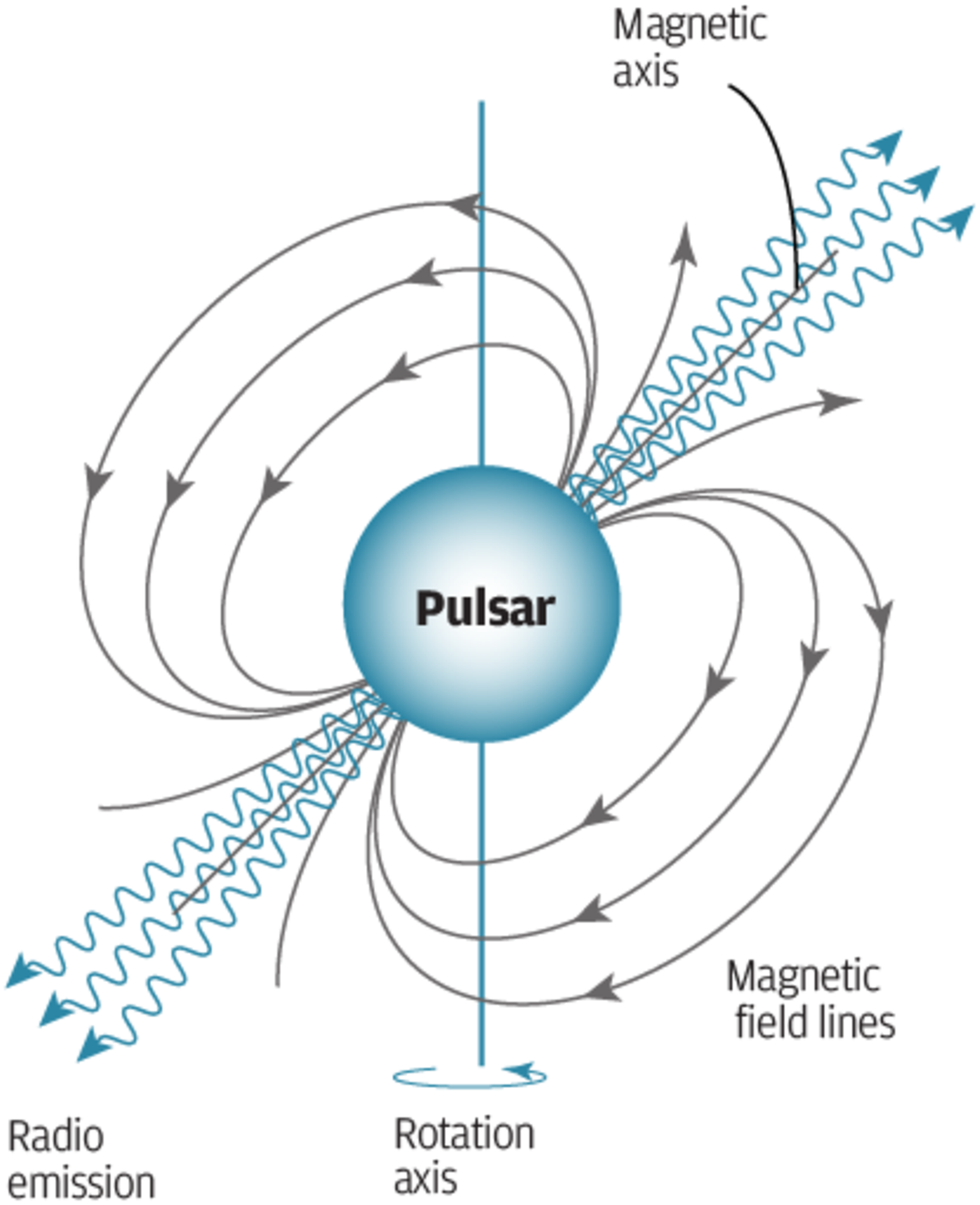}
\vspace{-0.5cm} 
\end{figure}

{\bf  Rotation Powered  Pulsars (RPP)  :} The  rotation powered  radio
pulsars are  so named  because of the  pulsed radio  emission received
from    them   (Fig.~\ref{f_pulsar}).     This   pulsed    emission,
characterized by its precise periodicity, is essentially {\em magnetic
  dipolar}  radiation that  comes  at the  expense  of the  rotational
energy. As a result the star  slows down.  Measuring this rate of slow
down ($\dot {P_s}$) we can estimate the surface magnetic field to be :
\beq
B = (\frac{3 I c^3 P_s \dot{P_s}}{8 \pi^2 R_s^6})^{1/2}
  \simeq   3.2\times  10^{19}   (P_s\dot{P_s})^{1/2}  \,   {\rm  G}\,,
  \nonumber
\eeq
where, $I$ ($\sim 10^{45} \mbox{ g cm}^{-2}$) is the moment of inertia
and $R_s$ is the radius of  the star.  Of course, this simple estimate
provides a  measure of the  long-range dipole field only,  measures of
higher multipoles  (known to exist near  the surface) etc. can  not be
obtained from this.

From the measurement of $\dot{P_s}$  we can also obtain an approximate
age  of a  radio  pulsar.  This  is known  as  the characteristic  age
($\tau_{\rm ch}$) and is given by,
\beq
\tau_{\rm ch} = \frac{P_s}{2\dot{P_s}}
\eeq
This estimate closely approximates the true age of a radio pulsar if -
a) the pulsar's initial spin-period  has been much smaller compared to
the  current observed  period,  b)  there has  been  no  decay of  the
magnetic field,  and c) the  energy loss  is due entirely  to magnetic
dipolar radiation.

The class of radio pulsars is again be divided into two distinct groups -
{\bf (a)} the classical radio  pulsars ({\bf PSR}) with $P \sim 1$~s,
$B \sim 10^{11} - 10^{13.5}$~G; \,\,
{\bf (b)} the millisecond radio  pulsars ({\bf MSRP}) with  $P \lsim
20$~ms, $B  \lsim 10^{10}$~G,  and having very  different evolutionary
histories.  (These are or have been  members of binary systems and are
believed to  have been `recycled' through  accretion episodes reducing
the value of both $P_s$ and $B_s$ in them.)

\begin{figure}
  \caption{A neutron star in an  HMXB. The accreting material is being
    channeled by the strong magnetic  field and the resulting emission
    is in the form of X-ray  pulses. Picture courtesy - {\em Dany Page,
      UNAM, Mexico City.}}
  \label{f_hmxb}
\vspace{-0.5cm}
\centering\includegraphics[width=10.0cm]{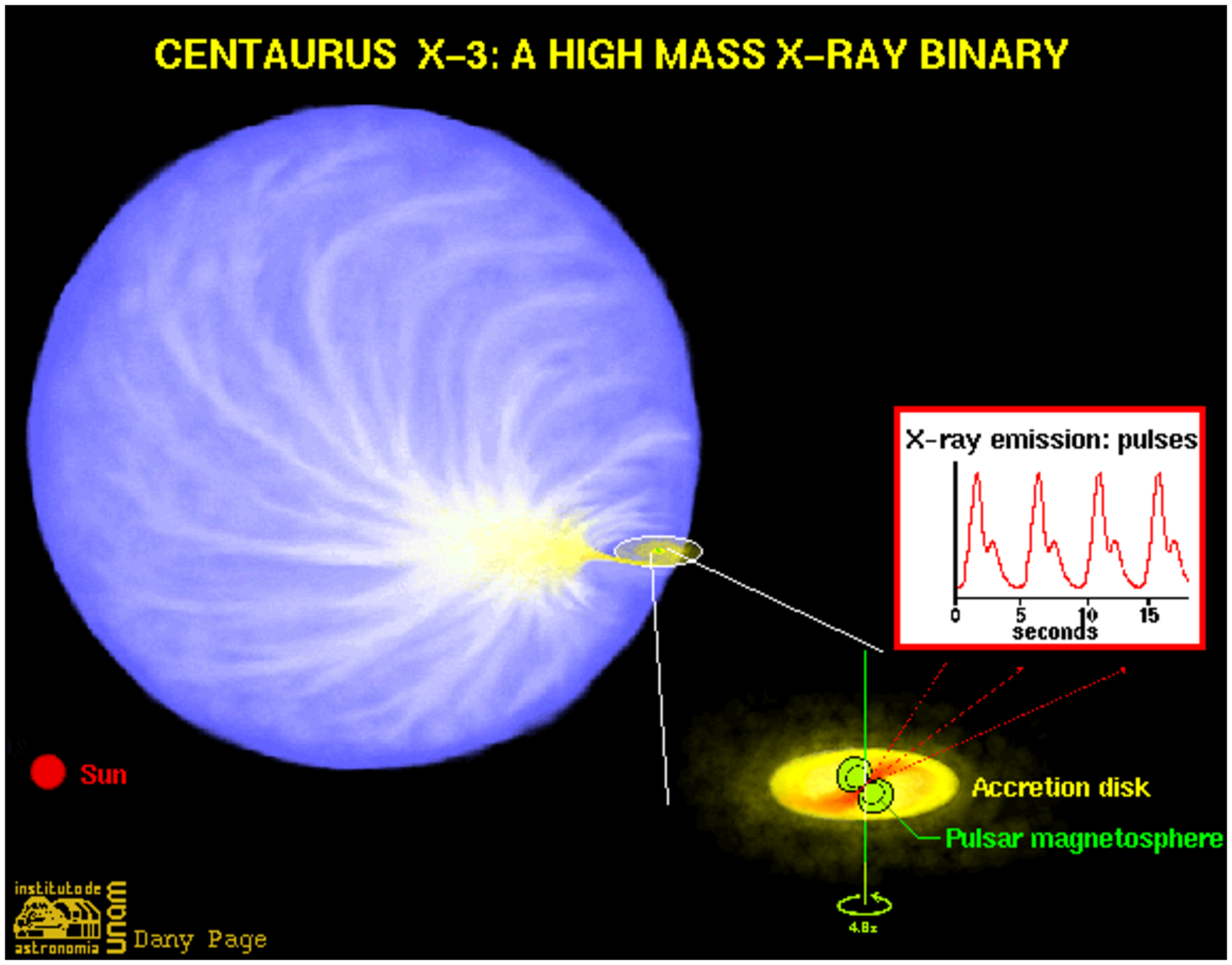}
\vspace{-0.5cm}
\end{figure}

\begin{figure}
  \caption{A neutron  star in  a low-mass  X-ray binary.  The accreted
    material  accumulates  on the  stellar  surface  and ignites  upon
    reaching appropriate temperatures. These sudden episodes of nuclear
    reaction are observed as X-ray bursts. Picture courtesy - {\em Dany
      Page, UNAM, Mexico City.}}
  \label{f_lmxb}
\vspace{-0.5cm}
\centering\includegraphics[width=10.0cm]{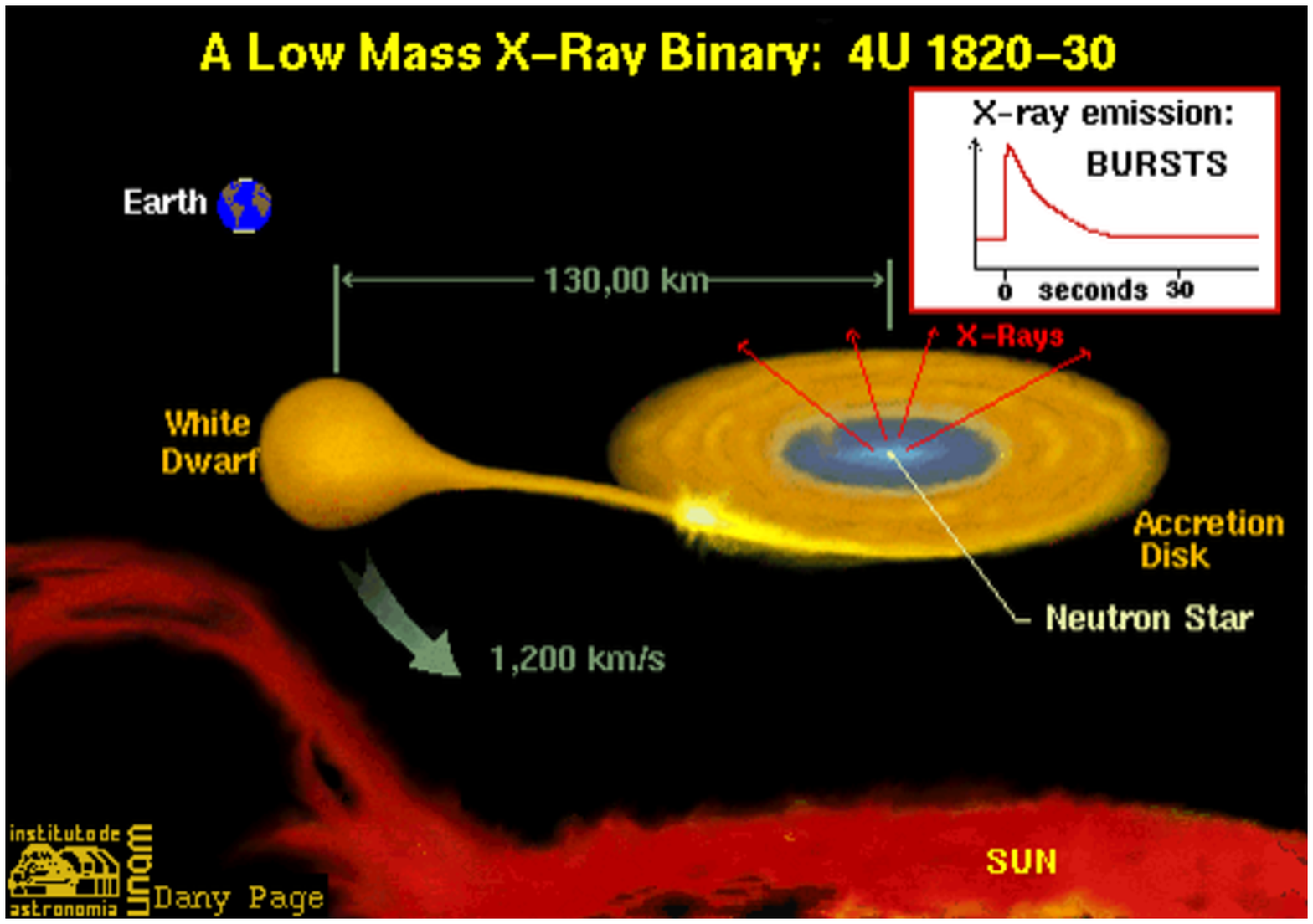}
\vspace{-0.5cm} 
\end{figure}

{\bf Accretion  Powered Pulsars (APP)  :} This class mainly  refers to
neutron  stars  in  binaries   that  are  undergoing  active  material
accretion from  the companion. Accretion  onto the neutron  star gives
rise to the energetic radiation in  APPs. Depending on the mass of the
donor star these are classified  as High-Mass X-ray Binaries (HMXB) or
Low-Mass    X-Ray    Binaries     (LMXB)    (Fig.~\ref{f_hmxb}    \&
Fig.~\ref{f_lmxb}).

Neutron stars in HMXBs typically have $B_p \sim 10^{12}$~G, and $O$ or
$B$ type stars  as companions. The strong magnetic  field channels the
flow of material from the companion to the surface of the star and the
emission  comes out  in the  form of  periodic pulses.   Consequently,
neutron  stars in  HMXBs mostly  show  up as  X-ray pulsars.   Photons
inside the  hot accretion column on  the surface of the  neutron stars
undergo  resonant   scattering  from  the  electrons,   generating  an
absorption  like  feature  at  the resonance  energies  of  the  final
emergent  spectra, known  as cyclotron  resonance scattering  features
(CRSF).   These  features are  used  to  estimate the  magnetic  field
strength of the  neutron star.  However, this estimate  only gives the
local field strength and uncertainties about local geometry prevent us
from  obtaining the  corresponding surface  dipole field  strength. So
field estimates obtained  from radio pulsars and  from X-ray binaries,
in   reality,  correspond   to  two   altogether  different   physical
quantities.

LMXBs, on the  other hand, harbour neutron stars  with magnetic fields
significantly weakened  ($B \lsim 10^{11}$~G) through  extended phases
of accretion.  Physical process taking place in such accreting systems
manifest   as   -   thermonuclear  X-ray   bursts;   accretion-powered
millisecond-period pulsations;  kilohertz quasi-periodic oscillations;
broad relativistic  iron lines;  and quiescent  emissions. It  is even
more difficult to measure the magnetic field of neutron stars residing
in LMXBs.  However, recent work on accreting millisecond X-ray pulsars
(where the accretion  rate could be low enough for  the neutron star's
weak magnetic field to be able to  channel such flow) have opened up a
new direction in understanding the physics of LMXBs containing neutron
stars.

{\bf Internal  Energy Powered Neutron Stars  (IENS) :} A mixed  bag of
objects, these IENS are so called  because the emission has its origin
in some internal mechanism or energy  source of the neutron stars. The
following are some of the prime examples of such objects.
   \ben
       \i {\bf Magnetars} are young,  isolated neutron stars and their
       emission  is  thought   to  be  due  to  the   decay  of  their
       super-strong magnetic fields.
       \i  The seven  isolated  neutron stars  ({\bf INS}),  popularly
       known as the {\em Magnificent Seven}, have blackbody-like X-ray
       spectra  ($T \sim  10^6$~K),  relatively nearby  and have  long
       spin-periods  ($P  \sim  -  10$~s).   They  are  probably  like
       ordinary pulsars but a combination of strong magnetic field and
       spatial proximity make them visible in the X-rays.
   \een

\section{Towards unification..}

One  of  the  important  questions  in neutron  star  research  is  to
understand  the connections  (or  absence  thereof) between  different
observational  classes.   There  are   indications  that  there  exist
evolutionary  links  between these  classes  and  the magnetic  field,
ranging  from  $10^8$~G  in  MSRPs to  $10^{15}$~G  in  magnetars,  is
instrumental in providing this link. It is expected that the evolution
of   the  magnetic   field  and   associated  phenomena   chart  those
evolutionary pathways.

\begin{figure}
  \caption{Different observational  classes of neutron stars  shown in
    the  spin-period vs.   surface magnetic  field ($P_s-B_s$)  plane.
    The blue and magenta arrows  indicate `recycling' of neutron stars
    in X-ray  binaries. The orange  and green arrows hint  at possible
    evolutionary pathways between high  magnetic radio pulsars and the
    magnetars.  The   `RPP  death-line'  corresponds  to   a  limiting
    combination of $P_s$ and $B_s$, on the left hand side of which the
    radio  pulsar mechanism  stops  working. The  `MSRP spin-up'  line
    indicates the maximum value of  $P_s$ that can be achieved through
    binary  processing for  a given  final value  of $B_s$.   {\bf \em
      Legends :}  I/B - isolated/binary,  GC - globular cluster,  GD -
    galactic disc,  EG -  extra-galactic objects.   See {\em  Konar et
      al., 2016, JOAA, 37, 36} to know about the data used here.}
  \label{f_bp}
\vspace{-0.5cm} 
\centering\includegraphics[width=10.0cm]{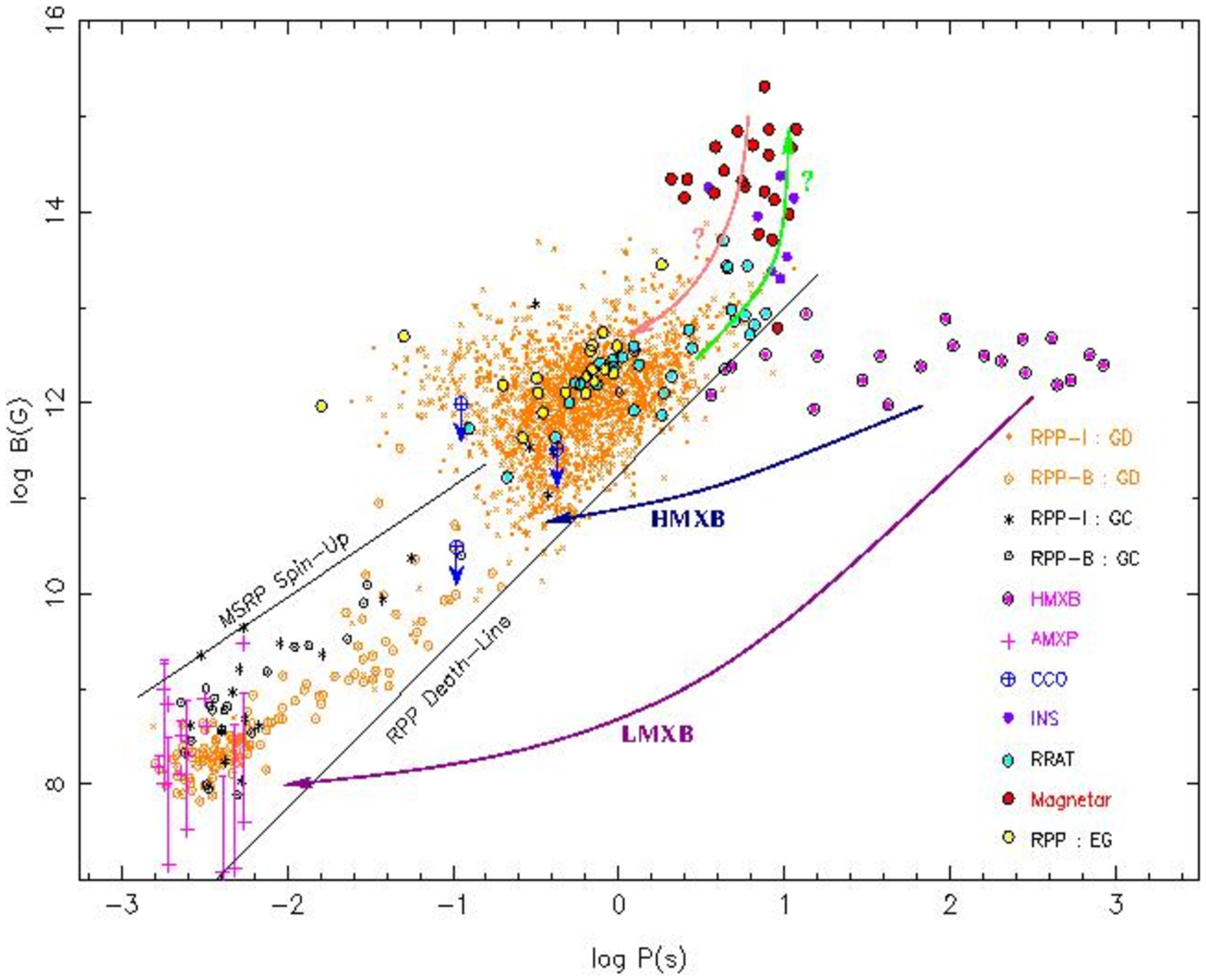}
\vspace{-0.5cm} 
\end{figure}
%


However, one of the major roadblocks encountered while developing a
theory of field evolution is the uncertainty about the process of
field generation, which determines the location of the currents that
support the field. Currently, three plausible theories are in vogue,
though each of them suffer from certain inherent inconsistencies.

The  simplest  theory  considers  the conservation  of  magnetic  flux
existing  inside the  core of  the progenitor  star through  supernova
explosion  and  core  collapse.   Flux conservation  would  ensure  an
increase in the field strength by a factor $(R_{\rm progenitor}/R_{\rm
  NS})^2$ ($\sim 10^{10}$) and magnetic  fields of strength $10^{12} -
10^{14}$~G can easily  be formed. Evidently, magnetic  field formed in
this  fashion would  be  located  in the  core  of  the neutron  star.
Another theory hold turbulent dynamo  processes, active in the core of
a neutron  star in the early  phases of its life,  responsible for the
generation of the magnetic field.

The magnetic field  can also be generated as a  consequence of thermal
effects occurring in the outer crust in the early phases of the star's
thermal evolution.  The field can grow  either in the liquid phase and
then be  convected into  the solid  regions, or it  could grow  in the
solid  crust itself.   The  coexistence  of a  heat  flux  and a  seed
magnetic field,  in excess of $10^8$  Gauss, in the liquid  will cause
the fluid to circulate which may lead to effective dynamo action which
would help the field grow stronger. The currents supporting this field
would be located in the outer crust of a neutron star.

Whatever the  origin, the  evolution of the  magnetic field  is either
spontaneous or happens as a consequence of material accretion.  Recent
investigations,  focusing on  the  magneto-rotational  evolution of  a
neutron in the early phases,  consider spontaneous evolution.  This is
important to understand  the evolution of isolated  neutron stars with
strong magnetic fields and may answer the question whether there exist
any evolutionary  link between the  magnetars and the  strong magnetic
field  radio pulsars.   On  the other  hand,  accretion induced  field
evolution  explains  the  processing  of ordinary  pulsars  (with  strong
magnetic fields and  long periods) that produce  MSRPs with ultra-fast
rotation  and  much  weaker   magnetic  fields.   All  these  pathways
(established or conjectured) have been summarised in Fig.~\ref{f_bp}.

It  goes without  saying that  a  lot of  questions remain  unanswered
regarding these disparate observational classes of neutron stars. With
the advent of finer technology and  bigger and better telescopes it is
expected that we would be in a better position to answer many of those
questions. However, detection of many  more neutron stars would likely
bring  many new  classes to  the  fore giving  rise  to a  new set  of
questions in the future.

\section*{Acknowledgment}

Dipankar Bhattacharya, my formal `guru' (thesis advisor), and G. Srinivasan
have not only taught me all that I have ever learned about neutron stars;
but have also managed to transmit their passions for these exotic objects
that challenge the very boundaries of known physics.


%

%

\end{document}